\documentclass[a4paper, 11pt]{article}

    \usepackage{graphicx}
    \usepackage{ subfig}
    \usepackage{color}
    \usepackage{bm}
    \usepackage{mathrsfs}
    \usepackage{ulem}

    \def\b{\begin{eqnarray}}
    \def\e{\end{eqnarray}}

    \def\n{\noindent}
\begin{document}

    \begin{center}
    {\huge \textbf{Dynamical Analysis of an $\bm{n}\,\,$--$\,\bm{H}\,$--$\,\bm{T}$  \vskip.23cm Cosmological Quintessence Real Gas \vskip.4cm Model with a General Equation of State}}

    \vspace {10mm}
    \noindent
    {\large \bf Rossen I. Ivanov and Emil M. Prodanov} \vskip.5cm
    {\it School of Mathematical Sciences, Technological University Dublin, \vskip.1cm Kevin Street Campus, Dublin 8,
    Ireland,} \vskip.1cm
    {\it E-Mails: rossen.ivanov@dit.ie, emil.prodanov@dit.ie} \\
    \vskip1cm
    \end{center}

    \vskip4cm
    \begin{abstract}
    \n
    The cosmological dynamics of a quintessence model based on real gas with general equation of state is presented within the framework of a three-dimensional dynamical system describing the time evolution of the number density, the Hubble parameter, and the temperature. Two global first integrals are found and examples for gas with virial expansion and van der Waals gas are presented. The van der Waals system is completely integrable. In addition to the unbounded trajectories, stemming from the presence of the conserved quantities, stable periodic solutions (closed orbits) also exist under certain conditions and these represent models of a cyclic Universe. The cyclic solutions exhibit regions characterised by inflation and deflation, while the open trajectories are characterised by inflation in a "fly-by" near an unstable critical point.

    \end{abstract}

    \newpage

    \section{Introduction}

    Supernova observations since 1998 have been showing that, over the last 5 billion years, the cosmic expansion has been accelerating \cite{acc, acc2}. That is, for a Friedmann-Robertson-Walker-Lema\^itre (FRWL) Universe, both the scale factor $a(t)$ and its time derivative $\dot{a}(t)$ have been increasing. This has been happening at the same rate, so that Hubble parameter $H = \dot{a} / a$ has been constant. In other words, in the standard cosmological model, the Universe is evolving towards exponential expansion $a(t) \sim \exp(Ht).$ The current epoch has been labeled as dark energy dominated era after un unknown component called dark energy which makes most of the energy content of the Universe and provides the negative pressure, needed to overpower the gravitational pull and explain the resulting accelerated expansion. For a review and list of references, see \cite{padm}. Some doubts have been recently cast on whether this is, indeed, the case \cite{lambda}. \\
    The cosmological principle --- on a very large scale, the distribution of matter in the Universe is homogeneous and isotropic --- most often leads to choosing  perfect fluids to model the matter and energy in the universe. These are defined as fluids that are isotropic in their rest frame. Associated with a perfect fluid, there is its equation of state which is a relationship between the pressure $p$ of the fluid and the energy density $\rho$. The vast majority of cosmological models are based on relationship of the type $p = \omega \rho$, where $\omega$ is a constant, independent on time. The Friedmann equation $\ddot{a}/{a} = - (4 \pi G / 3)(\rho + 3p)$ shows that, for cosmic acceleration, it is required that $\omega < -1/3$. On the other hand, the continuity equation (energy conservation equation) for the perfect fluid, $\dot{\rho} + 3H(\rho + p) = 0$ shows that $\rho + p$ must not be negative for a meaningful cosmological model in which the energy density of an expanding Universe decreases with time. This leads to $\omega \ge -1$. Dark energy could be defined as any physical field, which operates in the gap $-1 \le \omega < -1/3$ between the weak energy condition, $\rho \ge 0$ and $\rho + p \ge 0$, i.e. having positive energy density (to account for  the necessary density to make the universe flat) and realistic cosmology, and the strong energy condition, $\rho + p \ge 0$ and $\rho + 3 p \ge 0$, for which the part $\rho + 3 p \ge 0$ is violated to account for the needed negative pressure. There are many forms in which the dark energy is sought. For an extensive review and list of references, see \cite{noji}. Observationally favoured is the cosmological $\Lambda$ cold dark matter model (corresponding to $\omega = -1$), in which the role of dark energy is played by the cosmological constant $\Lambda$. Quintessence models are based on a dynamical, evolving, spatially-inhomogeneous component with negative pressure \cite{stein, caldw} and commonly considered quintessential cosmological models introduce a spatially-inhomogeneous slowly-evolving real scalar field rolling down a potential similar to the inflaton field in inflation theory. For a review and list of references, see \cite{quint}. Phantom cosmological models \cite{noji} violate all four energy conditions \cite{noj},  $\omega \ge - 1$, in particular. From quantum point of view, the phantom models are unstable, but it is not necessarily so from a classical perspective \cite{noj}. Such models exhibit Big Rip singularities (the scale factor $a(t)$ becoming singular over a finite time) and there are many proposed remedies for this --- see \cite{noj} and the references therein. Caldwell introduced \cite{caldwell2} the concept of phantom fields by constructing a toy model of a "phantom" energy component with $\omega < - 1$ and argued that it agrees, based on current data and understanding, with most classical tests of cosmology. Carroll {\it et al.} claim \cite{car} that that temporary violation of  $\omega \ge - 1$ is not incompatible with a well-defined model.  According to recent studies, see \cite{leandros} and the references therein, phantom cosmological models are preferred over quintessence ones.  \\
    The quintessential scheme can also be achieved without the usually discussed scalar fields. Alternative quintessence models introduce real gas equations of state \cite{cap1, cap2, cap3, ja, ivpro, ivpro2}. General Relativity with a perfect fluid of any type can be recast equivalently as a modified gravity theory \cite{od1, od11}. See \cite{od2, od22} for a general review. Attempting to reconcile General Relativity and the observed cosmic acceleration, even more exotic dark matter sources are sought. The dark energy equation of state has been generalised in many ways: allowing it to change its structure/form during the universe evolution \cite{ss}; introducing dark fluid with a time-dependent equation of state leading to multiple de Sitter space \cite{ss2}; considering non-linear inhomogeneous equations of state \cite{brev}; modifying the equation of state with an arbitrary function of the Hubble parameter $H$ and its derivatives \cite{fh}; abandoning the perfect fluid hypothesis \cite{card} and studying the Redlich--Kwong, the modified Berthelot, the Dieterici, and the Peng--Robinson real gasses resulting in $a(t)$  not diverging in any finite time so that any Big Rip is avoided even if $\omega$ may lie today in the phantom regime and others. \\
    The dark energy also provides for consistent models of cyclic cosmology, allowing the Universe to undergo infinitely many self-sustained cycles without failing the second law of thermodynamics, according to which the entropy can only increase, thus necessitating larger and larger successive cycles. The Steinhardt--Turok model \cite{st1, st2, st3, st4}, based on the ekpyrotic scenario and M-theory, demonstrates a Big Bang--Big Crunch sequence with entropy removal in each cycle. The Baum--Frampton model \cite{bf1, bf2} is based on phantom cosmology. There are many other studies on cyclic Universes --- see, for example,  \cite{ivpro, ivpro2, fh, hr}. \\
    In this study, a quintessence cosmological model, based on the most general real gas is presented as a nonlinear dynamical system of three variables --- the number density, the Hubble parameter and the temperature. The nonlinear dynamics is simplified by the existence of two global first integrals. In addition, there are special (second) integrals, defined and conserved on hyper-surfaces in the three-dimensional phase space. These surfaces are invariant manifolds, which separate the phase space into subspaces where certain types of dynamic behaviour takes place. \\
    Following a relatively general model description of real gas cosmology and introducing the three dynamical equations, a real gas with virial equation of state and a van der Waals gas are presented. For the latter, the system is completely integrable and the solutions are found. Various physically relevant possibilities for the critical points are identified and the corresponding dynamic behaviour is studied. Both periodic orbits and unbounded solutions are found in the model, depending on the initial conditions (and the level sets presented by the first integrals). The existing stable periodic solutions (closed orbits), which are models of a cyclic Universe, stem from the presence of the conserved quantities. Interestingly, periodic solutions exhibit regions characterised by inflation and deflation, while unbounded trajectories are characterised by inflation in a "fly-by" near an unstable critical point.

    \section{The Set-up}

    In this model, the Universe is presented classically as a two-component mixture of dust,  with energy density $\rho_d$ and pressure $p_d = 0$, and a most general real gas with equation of state expressing the real gas pressure  $p$ in terms of the real gas energy density $\rho$ and temperature $T$, i.e. $p = p(\rho, T)$. The energy density of the dust component, $\rho_d$, can be taken as positive (for example, one could think of the dust component as of ordinary baryonic matter in this case),  zero (absence of dust component), or, to reveal more mathematical aspects of the model, negative. Dust with negative energy density is not a new feature --- see \cite{d1, d2, d3, d4, d5, d6} and the references therein. \\
    The setting for the analysis of the two-fraction Universe is the Friedmann-Robertson-Walker-Lema\^itre (FRWL) cosmology with metric:
    \b
    ds^2 = g_{\mu \nu} dx^\mu dx^\nu =
    c^2 dt^2 - a^2(t) \Bigl[\frac{dr^2}{1 - kr^2} + r^2 (d \theta^2 + \sin^2 \theta \, d \phi^2) \Bigr],
    \e
    where $a(t)$ is the scale factor of the Universe and $k$ is the spatial curvature parameter. \\
    Einstein's equations are:
    \b
    G_{\mu \nu} + \Lambda g_{\mu \nu} = \kappa T_{\mu \nu},
    \e
    where $\kappa = 8 \pi G/c^4$  and the matter energy-momentum tensor $T_{\mu \nu}$, representing the two fractions of the Universe, collectively modelled with a perfect fluid, is given by:
    \b
    T_{\mu \nu} = (\tilde{\rho} + \tilde{p}) \, u_\mu \, u_\nu - \tilde{p} \, g_{\mu \nu} \, .
    \e
    Here $\tilde{\rho} = \rho_d + \rho$ and $ \tilde{p} = p$ are, respectively,  the cumulative density and pressure for both fractions and $u^\mu = dx^\mu / d \tau$ (with $\tau$ being the proper time) is the flow vector satisfying $g_{\mu \nu} u^\mu u^\nu = -1$. \\
    Friedmann equations are \cite{friedmann}:
    \b
    \label{fr1}
    \ddot{a} & = & - \frac{4 \pi G}{3} (\tilde{\rho} + \frac{3 \tilde{p} }{c^2})a + \frac{1}{3} \Lambda c^2 a  \\
    \label{fr2}
    \dot{a}^2 & = &  \frac{8 \pi G}{3} \tilde{\rho} a^2  + \frac{1}{3} \Lambda c^2 a^2 - c^2 k.
    \e
    Only the case of flat spatial three-sections ($k=0$) and without cosmological constant (i.e. $\Lambda = 0$) will be of interest. Also,  Planck units will be used, thus $\kappa= 1, c = 1, k_B = 1$. \\
    The Friedmann equations (\ref{fr1})--(\ref{fr2}) then become:
    \b
    \label{h1}
    \frac{\ddot{a}}{a} & = & - \frac{1}{6} (\rho_d + \rho + 3p), \\
    \label{h2}
    H^2 & = & \frac{1}{3} (\rho_d + \rho),
    \e
    where $H(t) = \dot{a}(t) / a(t)$ is the Hubble parameter. It will be one of the three dynamical variables of the model. As $\ddot{a} / a = \dot{H} + H^2$, combining the Friedmann equations allows to express:
    \b
    \label{dynami}
    \dot{H} =  - \frac{1}{2} (\rho_d + \rho + p),
    \e
    which will be considered as one of three dynamical equations. \\
    In absence of unbalanced particle creation or particle annihilation processes, the number of particles in the perfect  fluid is conserved and this is manifested, in locally flat coordinates, by the continuity equation:
    \b
    {\partial n \over \partial t}+\vec{\nabla}\cdot \left(n\vec{v}\right) = 0,
    \e
    where $n$ is the particle number density and the components of the velocity $\vec{v}$ are the spatial components of the four-velocity:
    \b
    u^\mu = (1 - v^2)^{-\frac{1}{2}}\,\,\Bigl(1,{d\vec{x}\over dt}\Bigr)=  (1 - v^2)^{-\frac{1}{2}}\,\,(1,\vec{v})
    = (1 - v^2)^{-\frac{1}{2}} \,\, \frac{dt}{d\tau}
    \e
    (in locally flat coordinates). \\
    Introducing the particle current $n^\mu = n \, u^\mu$, the particle conservation equation in covariant form can be written as $\nabla_\mu n^\mu = 0$ or
    \b
    \label{nn}
    \dot{n} + 3 H n = 0,
    \e
    that is, the co-moving number of particles, $n a^3$, is constant.
    The continuity equation for the real gas is:
    \b
    \label{h3}
    \dot{\rho} + 3H(\rho + p) = 0
    \e
    and that of the dust is:
    \b
    \label{h4}
    \dot{\rho_d} + 3H \rho_d  = 0.
    \e
    These continuity equations are energy conservation equations. \\
    The final equation is the equation of state. The usual cosmological models are based on a barotropic equation of state $p = \omega \rho$, where $\omega$ is a parameter which is a constant. If $\omega$ is in the range $-1 < \omega < - 1/3$, the model is called quintessence. Negative pressure is needed to achieve this. The cosmological constant or vacuum energy is modelled by $\omega = -1$. Models with $\omega < -1$ are also characterised by negative pressure and are called phantom field models. For the observed cosmic acceleration, it is required that $\omega$ must be smaller than $-1/3$ --- visible from the cosmic acceleration equation, (\ref{h1}). This amounts to $\rho + 3p < 0$ --- a part-violation of the strong energy condition ($\rho + p \ge 0$ and $\rho + 3 p \ge 0$). Dark energy could be defined as any physical field with positive energy density (to account for  the necessary density to make the universe flat) and negative pressure, violating the part $\rho + 3 p \ge 0$ of the strong energy condition. Phantom cosmological models, on the other hand, violate all four energy conditions, in particular, the weak energy condition: $\rho \ge 0$ and $\rho + p \ge 0$. The phantom field is unstable from a quantum field theory perspective, but  not necessarily so from the perspective of classical cosmology.  \\
    It the part $\rho + p \ge 0$ of the weak energy condition is violated, then, according to equation (\ref{h3}), an expanding Universe ($H = \dot{a} / a > 0$) would have growing density ($\dot{\rho} > 0$) and vice versa: a contracting universe will be characterised by diminishing density. As it will be shown, when $\rho_d$ is taken as negative in the presented model, there will be initial conditions which lead to quintessence closed trajectories (but not to phantom ones, namely, the projections of the closed orbits in three-dimensional $\rho$--$H$--$T$ phase space onto the $\rho$--$H$ plane will be convex rather than concave, which are associated with phantom cosmologies --- it is exactly over the "indented" region where the weak energy condition is temporarily violated). \\
    Not all of the above equations are independent. Differentiation of (\ref{h2}) with respect to time and substitution into it of: $\dot{H}$ from (\ref{h1}), $\dot{\rho}$ from (\ref{h3}), and $\dot{\rho_d}$ from (\ref{h4}), leads to an identity. \\
    Using equation (\ref{h2}), the dust energy density can be expressed as $\rho_d = 3 H^2 - \rho$ and eliminated from (\ref{dynami}) so that the following dynamical equation holds \cite{ivpro, ivpro2}:
    \b
    \label{hash}
    \dot{H} = - \frac{3}{2} H^2 - \frac{1}{2} p.
    \e
    To find the dynamical equation for the temperature of the Universe, consider the Gibbs equation \cite{maa,lima}:
    \b
    dS = \frac{1}{T} \,\, d \Bigl({\rho \over n}\Bigr) + \frac{p}{T} \,\, d \Bigl({1\over n}\Bigr) =
    -\left({\rho+p\over Tn^2}\right) \,\, dn
    + {1\over Tn} \,\, d\rho.
    \e
    Here $S$ is the specific entropy (entropy per particle) and, as it is a full differential, the following integrability condition must hold in thermodynamical variables $\rho$ and $n$:
    \b
    \biggl[ \frac{\partial}{\partial n}
    \biggl(\frac{\partial S}{\partial \rho}\biggr)_n\biggr]_\rho =
    \biggl[ \frac{\partial}{\partial \rho}
    \biggl(\frac{\partial S}{\partial n}\biggr)_\rho\biggr]_n
    \qquad
    \mbox{or} \qquad \biggl[\frac{\partial}{\partial n} \biggl(\frac{1}{Tn}\biggr)\biggr]_\rho =
    \biggl[\frac{\partial}{\partial \rho} \biggl(-\frac{\rho + p}{Tn^2}\biggr)\biggr]_n.
    \e
    Thus the integrability condition becomes
    \b
    n \biggl(\frac{\partial T}{\partial n}\biggr)_\rho +
    (\rho + p)\biggl(\frac{\partial T}{\partial \rho}\biggr)_n
    = T
    \biggl(\frac{\partial p}{\partial \rho}\biggr)_n.
    \e
    For any simple thermodynamical system, subject to the action of the generalised force $Z$, associated with the external parameter $\zeta$, the second initial proposition of thermodynamics leads to the existence of a thermic equation of state: $Z = Z(\zeta, T)$. Then, the following identity is valid: $(\partial Z / \partial \zeta)_T \,\, (\partial \zeta / \partial T)_Z  \,\, (\partial T / \partial Z)_\zeta = -1$. With the help of this, the integrability condition results in the following thermodynamic identity:
    \b
    \label{tdi}
    \rho + p = T \biggl( \frac{\partial p}{\partial T}\biggr)_n + n \biggl( \frac{\partial \rho}{\partial n}\biggr)_T.
    \e
    If the equation of state is substituted into this identity, a functional relationship between  $T,  \, \rho$ and $n$ stems. \\
    Using the number conservation equation (\ref{nn}) to express $n$ as $-\dot{n}/(3H)$ and the energy conservation equation (\ref{h3}) to exptess $\rho + p$ as $-\dot{\rho}/(3H)$, the following temperature evolution law is valid \cite{maa, lima}:
    \b
    \biggl(\frac{\partial T}{\partial n}\biggr)_\rho \dot{n}
    + \biggl(\frac{\partial T}{\partial \rho}\biggr)_n \,\, \dot{\rho}
     = - 3 H T
    \biggl( \frac{\partial p}{\partial \rho} \biggr)_n,
    \e
    namely:
    \b
    \label{T}
    \dot{T} = - 3 H T
    \biggl( \frac{\partial p}{\partial \rho} \biggr)_n.
    \e
    This will be the second dynamical equation of the system (the temperature $T$ being another dynamical variable). \\
    As third dynamical equation, the continuity equation (\ref{nn}) for the number density will be taken with the number density $n$ as the third dynamical variable. Therefore, the three-dimensional dynamical system is described by the equations:
    \b
    \label{edno'}
    \dot{n} \!\!\! & = & \!\!\! - 3 H n, \\ \nonumber \\
    \label{dve'}
    \dot{H} \!\!\! & = & \!\!\! - \frac{3}{2} H^2 - \frac{1}{2} p, \\
    \label{tri'}
    \dot{T} \!\!\! & = & \!\!\! - 3 H T \left( \frac{\partial p}{\partial \rho} \right)_n
    = - 3 H T \frac{\left( \frac{\partial p}{\partial T} \right)_n}{\Bigl( \frac{\partial \rho}{\partial T} \Bigr)_n} .
    \e
    In (\ref{dve'}), equation of state in the form $p = p(n, T)$ should be used as well as in (\ref{tri'}) for the determination of $(\partial p / \partial T)_n$. If, however, the equation of state is given as $p = p(\rho, T)$, then the mass density $\rho$ should be excluded with the help of the identity (\ref{tdi}) (the integrability condition) by expressing it as $\rho = \rho(n, T)$. From $\rho = \rho(n, T)$, one also calculates the derivative $(\partial \rho / \partial T)_n$ in the denominator in (\ref{tri'}).

    \section{Conserved Quantities and Global Behaviour. Virial Gas Example}

    There is a symmetry in the model: if one divides (\ref{tri'}) by (\ref{edno'}), an expression independent of $H$ stems:
    \b
    \label{simet}
    \frac{dT}{dn} = \frac{T}{n} \left( \frac{\partial p}{\partial \rho} \right)_n.
    \e
    Such, $H$-independent, first-order ordinary differential equation exists for the most general gas and leads to the existence of a first integral in the form $I = I(n, T)$, independent of $H$. \\
    In view of their respective continuity equations, (\ref{nn}) and ({\ref{h4}), the dust density $\rho_d$ and the number density $n$ are proportional: $\rho_d = C n$, where $C$ is a constant having the same sign as $\rho_d$ (to ensure positive number density $n$). \\
    In the absence of the dust component ($\rho_d = 0$), the trajectory in the phase space always lies on the hyper-surface $\rho = 3 H^2$, given by equation (\ref{h2}) with $\rho_d = 0$. Otherwise:
    \b
    \label{32}
    \rho_d = C n = 3H^2 - \rho(n, T).
    \e
    This constant $C$ is of paramount importance in the analysis --- it is the value of another first integrals present:
    \b
    \label{j}
    J(n, H, T) = \frac{3H^2 - \rho(n, T)}{n} = C = \mbox{ const.}
    \e
    (The fact that $J$ is a first integral can be easily seen by differentiating $J$ with respect to time and substituting the dynamical equations --- an identity will follow.) \\
    Absence of dust ($\rho_d = 0$) corresponds to $C = 0$. \\
    If the first integral $I(n,T) = $ const is also known globally, then the presence of two first integrals means that the system is completely integrable. \\
    To illustrate all this with an example, consider real gas whose pressure $p$ is related to the particle number $\mathcal{N}$, the temperature $T$, and the volume $V$ via the virial expansion:
     \b
     \label{vir}
     p = \frac{\mathcal{N} T}{V} \Bigl[ 1 + \frac{\mathcal{N}}{V} F(T) + \Bigl( \frac{\mathcal{N}}{V} \Bigr)^2 G(T) + \cdots \Bigr].
     \e
    The term $F(T)$ represents the first correction to the ideal gas equation of state ($p = \mathcal{N}T / V$). Interactions involving three (and more) particles are described by the second correction $G(T)$ (and the following terms) and will not be considered. The temperature $T_B$ at which $F(T)$ vanishes (then the real virial gas resembles ideal gas mostly) is called Boyle temperature. For temperatures below $T_B$, the correction term $F(T)$ is negative\footnote{The virial expansion for a gas, based on Lennard-Jones potential, breaks down not only at high densities, but also at low temperatures: it is divergent as $T \to 0$. This reflects the fact that the attractive interactions lower the energy of the gas constituents and at low temperatures there is a condensation into a liquid phase. (For the van der Waals gas, at temperature $T=0$, the pressure is finite: $p = - B n^2$). When a liquid is under tension, it pulls the confining surfaces. In the van der Waals case, for example, the liquid can withstand a maximum tension given by 27 times its critical pressure. Then the process of cavitation starts --- formation of gas bubbles and two phases (liquid and gas) co-exist. This is characteristic to systems with Lennard--Jones potential \cite{km}. The nucleation process can be homogeneous (for smooth and pure liquids) or not (for confined liquids or for liquids with impurities) and to trigger cavitation, the pressure in the fluid must fall below the saturated vapour pressure.  Gagnon and Lesgourgues \cite{f} consider $p_{DE} = \omega \rho_{DE}$ as the negative pressure that drives the cavitation process, where, in dark-energy domination, $-1 < \omega < - 1/3$ and the present-day dark energy density is $\rho_{DE} \sim 10^{-12} $ eV$^{4}$. In this case, the saturated vapour pressure is zero and, therefore, cavitation starts when the pressure becomes negative (which is the case if a dark energy component is present). The tensile strength of a fluid characterises its ability to counter-act the cavitation process: bubbles do not grow for as long as the force due to the intra-molecular interactions balances the outward pressure \cite{f}. The tensile strength of a fluid is equal to the absolute value of the minimum negative pressure which the fluid can sustain without breaking apart and to achieve dark energy domination, while still maintaining a valid fluid description (the more bubbles form, the more the hydrodynamic description fails), the pressure should be negative but its absolute value should not grow above the tensile strength of the fluid \cite{f}. For this reason, real virial gas description would not be applicable below a certain (very low) temperature, or the term $F(T)$ will have to be regularised.}. \\
    For example, when $F(T) = A - B/T$, where $A$ and $B$ are two positive constants (with $A$ measuring the strength of the attractive force between the gas ingredients and $B$ being the volume per mole of substance), the van der Waals equation of state results. \\
    In terms of the number density $n$, the virial equation of state is:
    \b
    \label{eos}
    p (n, T) = n T  [1 + n F(T)].
    \e
    Assuming $n$ and $T$ as the independent thermodynamical variables, finding $(\partial p / \partial T)_n$ from the above and substituting it, together with $p$, into (\ref{tdi}), results in the following differential equation:
    \b
    \biggr[ \frac{\partial}{\partial n} \biggl( \frac{\rho}{n} \biggr) \biggr]_T = - T^2 F'(T).
    \e
    This integrates directly into:
    \b
    \rho = n [ \phi(T) - n T^2 F'(T) ],
    \e
    where $\phi(T)$ can be determined as follows. One can take an ideal gas limit, i.e. set the co-efficient $F(T)$ of the second term of the virial expansion to zero. If this gas is monoatomic and has three translational degrees of freedom, the average kinetic energy of the particles is $(3/2) T$. Also, $n = (\mathcal{N} m) / (V m) = (M / V) (1/m) = \rho / m$, where $M$ is the mass of the system and $m$ is the relativistic mass of a representative particle: $m = m_0 + (1/2) m_0 u^2 + O(u^4).$ Here $m_0$ is the rest mass and $u$ --- the speed of the particle. Thus, the mass density of an ideal gas can be approximately written as $\rho = n [m_0 + (3/2) T]$. Therefore, $\phi(T) = m_0 + (3/2) T$. The number density $n$, the mass density $\rho$ and the temperature $T$ of a real virial gas are related as follows:
    \b
    \label{rvg}
    \rho = n (m_0 + \frac{3}{2} T) - n^2 T^2 F'(T).
    \e
    Thus the dynamical equations for the virial gas become:
    \b
    \label{edno''}
    \dot{n} \!\!\! & = & \!\!\! - 3 H n, \\
    \label{dve''}
    \dot{H} \!\!\! & = & \!\!\! - \frac{3}{2} H^2 - \frac{1}{2} nT - \frac{1}{2} n^2 T F(T), \\
    \label{tri''}
    \dot{T} \!\!\! & = & \!\!\! - 3 H T \,\, \frac{1 + n [F(T) + T F'(T)]}{\frac{3}{2} - n T[2F'(T) + T F''(T)]}.
    \e
    One also has to make sure that the mass density $\rho$ is not negative. This sets a range of allowed values of the number density $n$ as a function of the temperature $T$. In the case of real virial gas,  requesting $\rho \ge 0$ in equation (\ref{rvg}), leads to $n \le [ m_0 + (3/2) T ] / [T^2 F'(T)]$ for positive $F'(T)$. If $F'(T)$ is negative, then $\rho > 0$ is always satisfied.  In other words, one either has to have: (i) a "standard" virial gas, for example, a van der Waals gas or gas based on Lennard--Jones type of potential, for which $F(T) = \alpha - \beta e^{\epsilon/T}$ and $F'(T) > 0$ always ($\alpha, \, \beta,$ and $\epsilon$ are all positive constants), combined with an upper limit on $n$ (which conforms with the virial expansion over the powers of small $n$), or (ii) an anomalous virial gas for which $F(T)$ is a function monotonically decreasing with the temperature  ($n$ does not have to have an upper limit in this case). \\
    For the real virial gas one has:
    \b
    \label{weak}
    \rho + p = n \left[ m_0 + \frac{5}{2}T + n T [F(T) - T F'(T)] \right].
    \e
    As it has been made sure that $n$ is positive, one can keep track of when $\rho + p \ge 0$ (part of the weak energy condition) --- it will be satisfied for as long as:
    \b
    F(T) - T F'(T) \ge -\frac{m_0 + \frac{5}{2}T}{nT}.
    \e
    For example, for the van der Waals gas, the mass density will be positive for as long as $n \le [ m_0 + (3/2) T ] / B$. For a gas of electrons ($m_0 =  511$ keV or $10^9$ K approximately), for quite high temperatures $T$, the term $T/m_0$ will be quite small and negligible and the upper limit on the number density could be considered as $m_0/B$. This is well above what is assumed for $n$ as the virial expansion is over the powers of small $n$. On the other hand, for a van der Waals gas, $\rho + p \ge 0$ will be satisfied for temperatures  $T$ such that $T \ge (2 B n - m_0) / ( An + 5/2)$. If the density $n$ is below $m_0/(2B)$, which is quite big, then $\rho + p \ge 0$ will always be satisfied. \\
    The first integral $J(n,H,T)$  for the case of a virial gas is:
    \b
    \label{i2}
    J_{vir}(n, H, T) = \frac{3H^2}{n}  -  m_0 - \frac{3}{2} T + n T^2 F'(T)  = C  \mbox{ = const.}
    \e
    It follows from (\ref{32}) that $n = 0$ leads to $H=0$. The plane $n = 0$ is tangent to the hyper-surface $J(n, H, T) = C$. This hyper-surface will
    be entirely above the plane $T = -(2/3)(C + m_0)$, as point $[n = 0, H = 0, T =  -(2/3)(C + m_0)]$ is its absolute minimum. For $C = - m_0$, the plane $T = 0$ is tangent to the hyper-surface. For $C > - m_0$, the plane $T = 0$ crosses the hyper-surface. Thus, the origin necessarily lies on the hyper-surface for any $C \ge - m_0$. As already mentioned, the signs of $\rho_d$ and $C$ are the same, while absence of dust ($\rho_d = 0$) corresponds to $C = 0$. As this case serves as a natural separator, it will be useful to introduce the separatrix hyper-surface $J(n, H, T) = 0$ (see Figure 1a). The absolute minimum of this separatrix is at $T = -2 m_0 / 3$. Thus, if a hyper-surface is above this separatrix, it will correspond to a negative $\rho_d$, while if it is below --- to a positive $\rho_d$.

    \begin{figure}[!ht]
    \centering
    \subfloat[\scriptsize The first integral $J_{vdw}(n, H, T) = 3H^2 / n  -  m_0 - (3/2) T + B n  = C  \mbox{ = const}$ for different initial conditions. The separatrix ($C = 0$) corresponds to absence of dust.]
    {\label{F1}\includegraphics[height=6.5cm, width=0.47\textwidth]{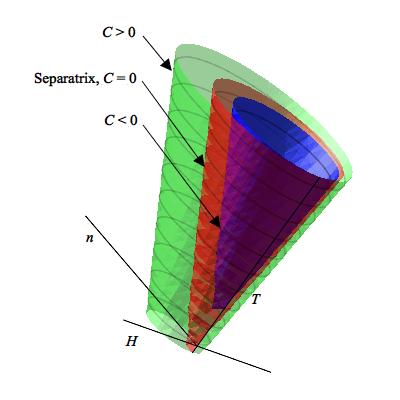}} \qquad
    \subfloat[\scriptsize The other first integral $I_{vdw}(n,T) = T\,n^{-2/3} e^{-2 A n/3} = D \mbox{ = const } > 0$ for different initial conditions.]
    {\label{F2}\includegraphics[height=6.5cm, width=0.47\textwidth]{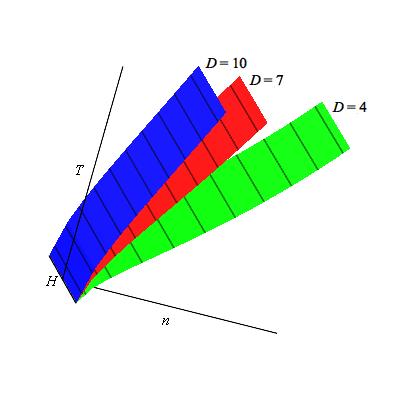}}
    \caption{\footnotesize{The completely integrable van der Waals system.}}
    \label{figure1}
    \end{figure}

    \section{The Completely Integrable Van der Waals System}

    \begin{figure}[!ht]
    \centering
    \subfloat[\scriptsize The "cone" given by the first integral $J_{vdw}(n, H, T)  = C > - m_0$, intersected with the other first integral $I_{vdw}(n,T) = D$ . The tip of the "cone" is at $T < 0$. The trajectories are open curves passing through the origin. Once a trajectory gets into the origin, it ends there.]{\label{F3}\includegraphics[height=4.4cm, width=0.24\textwidth]{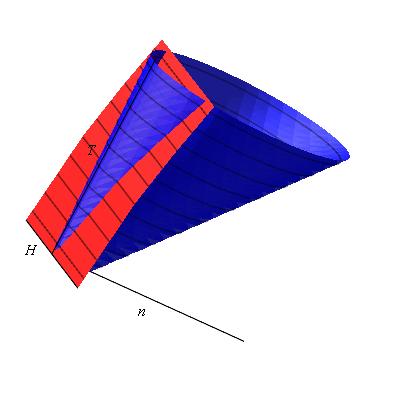}}
    \,
    \subfloat[\scriptsize Same as in (a), but with a less-rapidly growing $I_{vdw}(n,T) = D$ --- corresponding to a value of $D$ smaller than the one in (a). The trajectories are either open curves, not passing through the origin, or closed curves containing the origin and ending there.]{\label{F4}\includegraphics[height=4.4cm,width=0.24\textwidth]{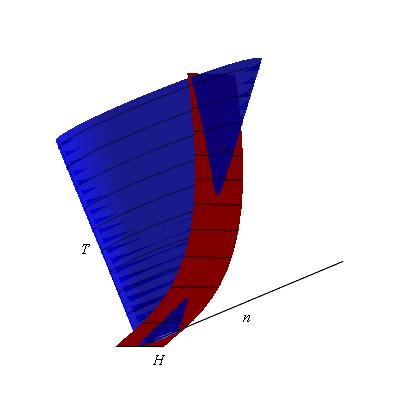}}
    \,
    \subfloat[\scriptsize The tip of the "cone" $J_{vdw}(n, H, T)  = C < - m_0$ is above the $T=0$ plane. The trajectories are open curves which do not pass through the origin.]{\label{F8}\includegraphics[height=4.4cm,width=0.24\textwidth]{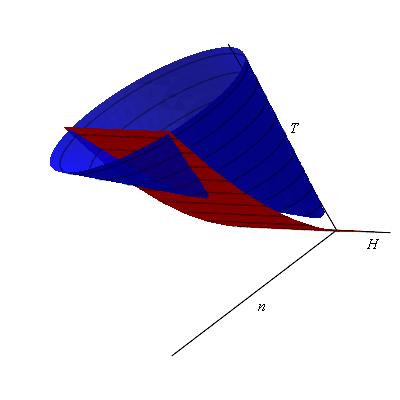}}
    \,
    \subfloat[\scriptsize Same as in (c), but, again, with a less-rapidly growing $I_{vdw}(n,T) = D$, i.e. smaller $D$. The trajectories are either open curves or closed curves, not passing through the origin.]{\label{F9}\includegraphics[height=4.4cm,width=0.24\textwidth]{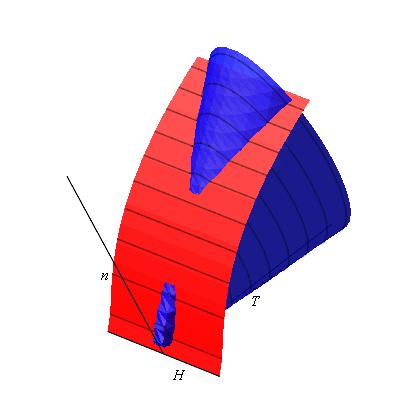}} \\
    \subfloat[\scriptsize $J_{vdw}(n, H, T)  = C = - m_0$. The tip of the "cone" is on the $T=0$ plane. The trajectories are open curves not containing the origin.]{\label{F5}\includegraphics[height=4.5cm, width=0.28\textwidth]{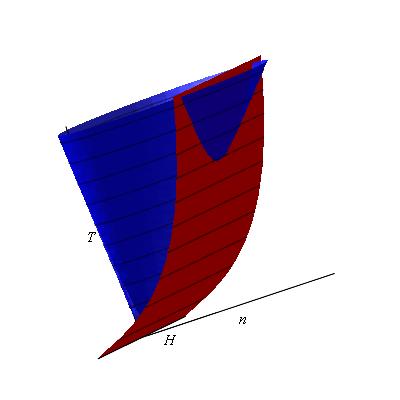}}
    \qquad
    \subfloat[\scriptsize Same as in (e), but with a more-rapidly growing $I_{vdw}(n,T) = D$. The trajectories are open curves extinguishing in the origin.]{\label{F6}\includegraphics[height=4.5cm, width=0.28\textwidth]{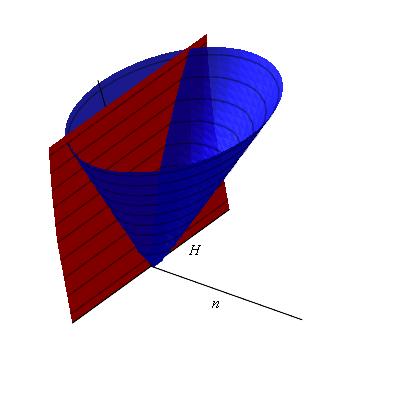}}
    \qquad
    \subfloat[\scriptsize Same as in (e), but with a more-rapidly growing $I_{vdw}(n,T) = D$, while not as fast as in (f). The trajectories are either closed curves extinguishing in the origin or open curves that do not contain the origin.]{\label{F7}\includegraphics[height=4.5cm, width=0.28\textwidth]{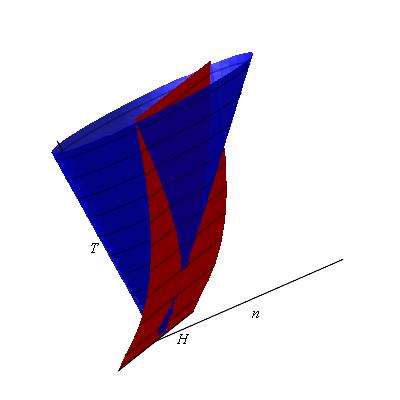}}
    \caption{\footnotesize{The trajectories in the van der Waals phase space are the intersections of the two hyper-surfaces given by the two first integrals
    $J_{vdw}(n, H, T) = 3H^2 / n  -  m_0 - (3/2) T + B n  = C  \mbox{ = const}$ and $I_{vdw}(n,T) = T\,n^{-2/3} e^{-2 A n/3} = D \mbox{ = const } > 0.$}}
    \label{figure2}
    \end{figure}

    \noindent
    In the case of a van der Waals gas, the first integral $J$ is simply (see Figure 1b):
    \b
    \label{i2-vdw}
    J_{vdw}(n, H, T) = \frac{3H^2}{n}  -  m_0 - \frac{3}{2} T + B n  = C  \mbox{ = const.}
    \e
    For a van der Waals gas, the first integral $I(n,T)$ can be obtained explicitly in a straight-forward manner: using $F(T) = A - B/T$ and dividing (\ref{tri'}) by (\ref{edno'}) yields the ordinary differential equation in separate variables:
    \b
    \label{tova}
    \frac{dT}{dn} = 2T \, \frac{1+An}{3n}.
    \e
    The solution of (\ref{tova}) is:
    \b
    T = D n^{2/3} e^{2 A n/3},
    \e
    where $D$ is a positive constant, equal to the value of the other first integral for the van der Waals gas (see Figure 1b):
    \b
    \label{eta}
    I_{vdw}(n,T) = T\,n^{-2/3} e^{-2 A n/3} = D \mbox{ = const } > 0.
    \e
    Therefore, the van der Waals system is completely integrable. \\
    Obviously, $n=0$ leads to $T=0$ for all $H$, including $H = 0$. Thus, the hyper-surface $I(n, T) = D$ contains the origin for any value of the constant $D$. \\
    The trajectories in the van der Waals phase space are obtained as intersections of the two hyper-surfaces given by the two first integrals $J_{vdw}(n, H, T) = 3H^2 / n  -  m_0 - (3/2) T + B n  = C  \mbox{ = const}$ and $I_{vdw}(n,T) = T\,n^{-2/3} e^{-2 A n/3} = D \mbox{ = const } > 0$. All possible cases are given on Figure 2 (see also Figures 3 and 4). \\
    From the first integrals, one can immediately find the solutions:
    \b
    T(n) & = & D n^{2/3} e^{2 A n/3}, \\
    H(n) & = & \pm \sqrt{\frac{1}{3}(C + m_0) n + \frac{1}{2} D n^{5/3} e^{2 A n/3} - \frac{1}{3} B n^2 },
    \e
    where $n(t)$ is determined by separation of variables from $\dot{n} = -3Hn$:
    \b
    \label{solu}
    \int \frac{dn}{n \sqrt{\frac{1}{3}(C + m_0) n + \frac{1}{2} D n^{5/3} e^{2 A n/3} - \frac{1}{3} B n^2 }} \,\, = \,\, \mp \,3(t - t_0).
    \e
    For small $n$, one finds easily that
    \b
    \label{assy}
    n(t) \simeq \frac{4}{3(C+m_0)(t-t_0)^2}
    \e
    --- behaviour similar to that in the $T = 0$ plane. The origin is reachable in infinite time. \\
    Formula (\ref{assy}) is valid for $C > - m_0$. When $C < - m_0$, the trajectory is either an open curve (Figures 2c and 2d) or a periodic one (Figure 2d), but, in either case, never over regions where $n$ approaches zero asymptotically, thus $C < - m_0$ is not applicable for small $n$. In the limiting (separatrix) case $C= -m_0$ and for small $n$, the relevant situations are the ones given by the open trajectories on Figure 2f and by the closed curves, terminating at the origin, on Figure 2g. Then one has
    $ n(t) \sim (t-t_0)^{-6/5}$ and, qualitatively, the same asymptotic behaviour --- with a negative power of $t$ and trajectory extinguishing in the origin in infinite time. \\
    As it is usually done, one can extend the validity of the model by allowing the consideration of large number densities. For the unbounded van der Waals trajectories and for large values of $n$ (see Figure 2 and Figure 3b), a blow-up is observed in finite time. In order to find the asymptotic behaviour near the time of the blow-up, retain the leading terms in (\ref{solu}). This yields the equation
    \b
    \sqrt{\frac{2}{D}}\int n^{-11/6} e^{-An/3}dn=-3 \sigma (t-t^*),
    \e
    where $\sigma = \mathrm{sign}(H)$ and $t^*$ is an integration constant. \\
    The integral on the left-hand side has, for $n\to \infty$, the asymptotic behaviour \linebreak $-\frac{3}{A}n^{-11/6}\exp(-\frac{An}{3}),$ thus
    \b
    n^{-11/6} e^{-An/3}=A\sqrt{\frac{D}{2}} \sigma (t-t^*).
    \e
    When $n\to \infty$, the left-hand side approaches zero and hence $t\to t^*.$ Therefore, $t^*$ is the blow-up time and the above formulae are valid for $t<t^*$ only. This is clearly possible only when $\sigma =-1$ and $H<0$. Hence $H\to -\infty $ and this blow-up represents a Big Crunch:
    \b
    n(t) \simeq -\frac{3}{A}\ln |t-t^*|.
    \e
    The integration in (\ref{solu}) cannot be performed explicitly in the general case, but the qualitative behavior can be inferred from the previous (and the following) analysis.

    \begin{figure}[!ht]
    \centering
    \subfloat[\scriptsize Closed trajectories corresponding to $C < - m_0$ (not containing the origin). All curves start at $n_0 = 1$ and $H_0 = 0$, but the initial temperature varies: $T_0 = 30$ (for the outer-most), $25, 20, 15, 10$.]{\label{F10}\includegraphics[height=4.4cm, width=0.315\textwidth]{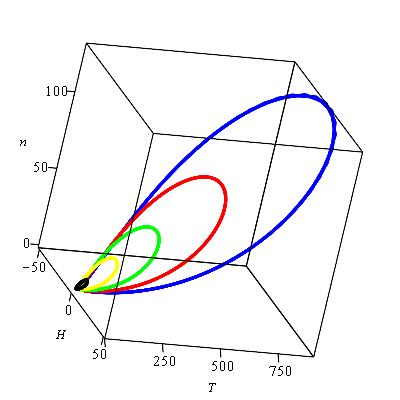}}
    \quad
    \subfloat[\scriptsize Open trajectories diverging to a Big Crunch. The trajectories start at $n_0 = 1$ and $H_0 = 1$. The initial temperatures are $T_0 = 40, 50, 60, 70$ (bottom to top).]{\label{F21}\includegraphics[height=4.4cm,width=0.315\textwidth]{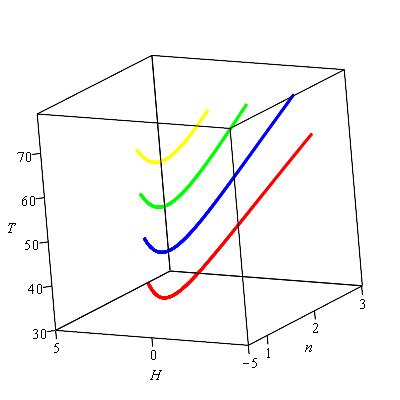}}
    \quad
    \subfloat[\scriptsize  Trajectoreis, parts of closed curves, extinguishing at the origin. These correspond to $C > - m_0$. The initial conditions are $n_0 = 30$, $H_0 = -1$, and $T_0 = 4, 11, 19, 27, 35$ (bottom to top).]{\label{F22}\includegraphics[height=4.4cm,width=0.315\textwidth]{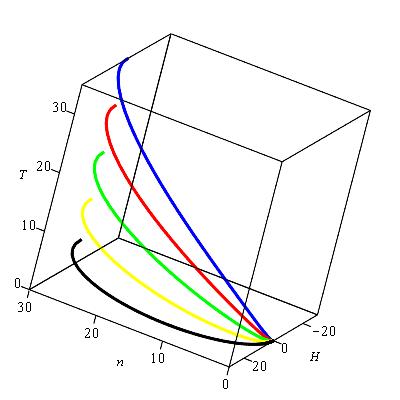}}
    \caption{\footnotesize{Influence of different initial conditions on the trajectories in the van der Waals phase space. The gas parameters are taken as $A = 10^{-3}$ and $B = 10$ (smaller $A$ corresponds to weakly interacting gas constituents, while larger $B$ means bigger volume per mole of substance). The Boyle temperature is $T_B = B/A = 10^{4}$. The dynamical system has been solved numerically with Maple.}}
    \label{figure3}
    \end{figure}

    \begin{figure}[!ht]
    \centering
    \subfloat[\scriptsize Projection of the trajectory onto the $n$--$T$ plane.]{\label{F11}\includegraphics[height=4.3cm, width=0.3\textwidth]{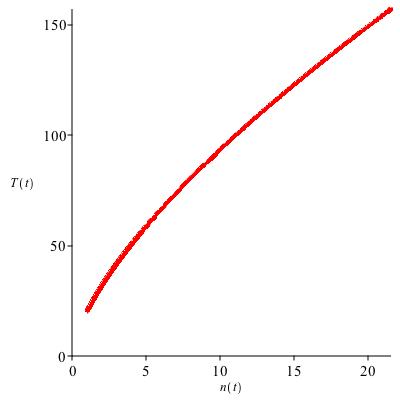}} \quad
    \subfloat[\scriptsize Projection of the trajectory onto the $n$--$H$ plane.]{\label{F12}\includegraphics[height=4.3cm,width=0.3\textwidth]{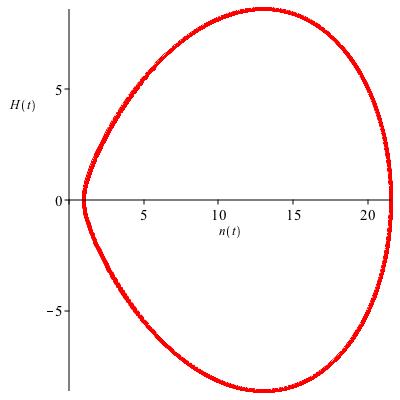}}
    \quad
    \subfloat[\scriptsize Projection of the trajectory onto the $H$--$T$ plane.]{\label{F13}\includegraphics[height=4.3cm,width=0.3\textwidth]{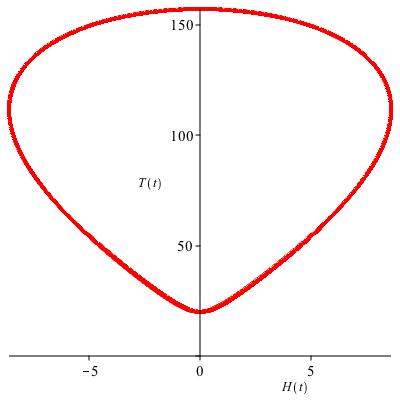}} \\
    \quad
    \subfloat[\scriptsize Dependence of the number density on time.]{\label{F14}\includegraphics[height=4.3cm,width=0.3\textwidth]{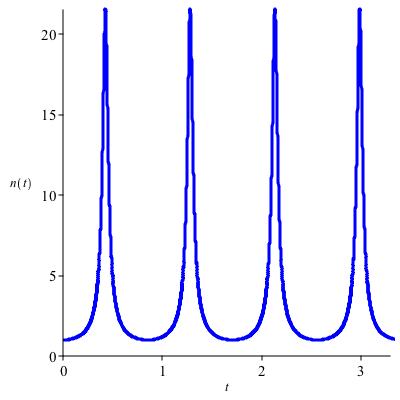}}
    \quad
    \subfloat[\scriptsize Dependence of the Hubble parameter on time. ]{\label{F15}\includegraphics[height=4.3cm, width=0.3\textwidth]{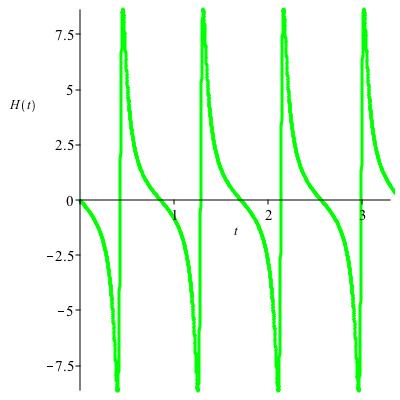}}
    \quad
    \subfloat[\scriptsize Dependence of the temperature on time.]{\label{F16}\includegraphics[height=4.3cm, width=0.3\textwidth]{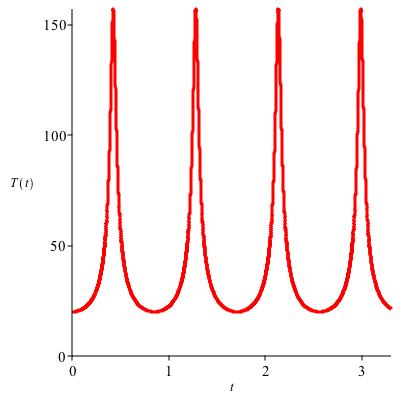}}
    \caption{\footnotesize{Analysis of the van der Waals closed trajectory given by the middle curve in Figure 3(a). This corresponds to $C < - m_0$. The initial conditions are: $n_0 = 1$, $H_0 = 0$, and $T_0 = 20$.}}
    \label{figure4}
    \end{figure}

    \section{Critical Points and Second Integrals}

    At an equilibrium point of the general dynamical system  (\ref{edno'})--(\ref{tri'}), all time derivatives on the left-hand sides of the equations vanish simultaneously. \\
    Clearly, if $H = H^* = 0$ and if the pressure vanishes, i.e. $p(n^*, T^*) = 0$, then the set $\{ [ n^*, H^* = 0, T^*(n^*) ] \}$ is a continuum of equilibrium points. Separately, if $p(n, 0) = 0$, then there is another continuum of equilibrium points: the set $\{ [ n^{**}, H^{**} = 0, T^{**}=0 ] \}$. Finally, if $p(0, T) = 0$, then there is a third continuum of equilibrium points: the set $\{ [ n^{***}=0, H^{***} = 0, T^{***} ] \}$. \\
    As mentioned earlier, the system will choose one equilibrium point from the curve $T^*(n^*)$. That is, there is another equation which, together with $p(n^*, T^*) = 0$, determines $n^*$ and $T^*$. This can be seen as follows. $I (n, T)$ is constant, that is, for the initial condition $(n_0, T_0)$ at initial time $t = t_0, \,\,$  $I(n, T) = I(n_0, T_0) = I_1 = $ const. On the other hand, the curve $T^* = T^*(n^*)$ with the equilibrium points (Figure 5) intersects $I(n, T)$ exactly at point with coordinates $(n^\ast, T^\ast)$, namely, the equation $I(n^\ast, T^\ast) =  I(n_0, T_0) = I_1 = $ const, together with the equation $p(n^\ast, T^*) = 0$ are the two simultaneous equations giving the pair $n^\ast$ and $T^\ast$ --- as dependent on the choice of initial conditions. \\
    Not all trajectories in the phase space are affected by the equilibrium points $(n^\ast, H^*= 0, T^\ast)$. This is due to the presence of second integrals in the phase space which fragment the phase space by non-crossable "walls" with each piece being a centre manifold. A second integral $K(\vec{x}) = 0$ of an autonomous dynamical system of the type $\dot{\vec{x}}(t) = \vec{f}[\vec{x}(t)]$ is an invariant, but only on a restricted subset, given by its zero level set \cite{gor}. It is defined by $\dot{K}(\vec{x}) = \mu(\vec{x}) K(\vec{x})$. Second integrals reduce to first integrals when $\mu = 0$ and to time-dependent first integrals when $\mu = $ const \cite{gor}. The surface $K_1$, defined by $3H^2 - \rho(n, T) = 0$, is a second integral and an invariant manifold, i.e. all trajectories originating from this invariant manifold remain there, and no trajectories originating from outside can penetrate this manifold, i.e. no trajectory can cross $3H^2 - \rho(n, T) = 0$. To see that the surface $K_1$ is a second integral, differentiate with respect to time: $(d/dt) [3H^2 - \rho(n, T)] = 6H \dot{H} - \dot{\rho} =   -3H [3H^2 - \rho(n,T)]$, as can be seen from the equations of motion. The hyper-surface $3H^2 - \rho(n, T) = 0$ corresponds to absence of dust, i.e. $C = 0$ [this is the Friedmann equation (\ref{h2}) with $\rho_d = 0$, namely, $C = 0$]. Therefore, in the absence of dust, there will be a relationship between $n$, $H$, and $T$, stemming from $3H^2 - \rho(n, T) =0$. In view of this, only two of the three dynamical equations (\ref{edno'})--(\ref{tri'}) are independent and the motion in the three-dimensional phase space is along the hyper-surface with equation $3H^2 - \rho(n, T) =0$. \\
    No trajectory can cross the parabolic wall given by $K_1$. This can be seen from a physical point of view too --- during the evolution, the dust component cannot disappear (or change the sign of $\rho_d$). \\
    The surface $K_2$, defined by $n=0$, is another invariant manifold, i.e. $n=0$ is another second integral. For example, for the virial gas, the motion on the  $K_2$ surface is given by $\dot{H} = -(3/2)H^2$ and $\dot{T} = -2HT$. These equations integrate easily and the solutions for initial data $(H_0,T_0)$ at $t=t_0$ are
    \b
    \label{n0a}
    H(t) & = & \frac{H_0}
    {1 + \frac{3H_0}{2} (t - t_0)}, \\
    \label{n0b}
    T(t) & = & \frac{T_0}
    {\left[1 + \frac{3H_0}{2} (t - t_0)\right]^{\frac{4}{3}}}.
    \e
    Thus, the long-time asymptotic is  $(H,T)\to(0,0).$  The trajectories in $n=0$ converge to the origin over infinite time (bearing in mind that the origin is a critical point too). As the $T$ axis is a continuum of equilibrium points, should the equation of state satisfies $p(n,0) = 0$, then the critical points $(0, 0, T^{***})$ are not reachable due to the second integral $K_2$. Thus, a trajectory cannot end anywhere on the invariant plane $n=0$, except at the origin.

    \begin{figure}[!ht]
    \centering
    \includegraphics[height=4.3cm, width=0.4\textwidth]{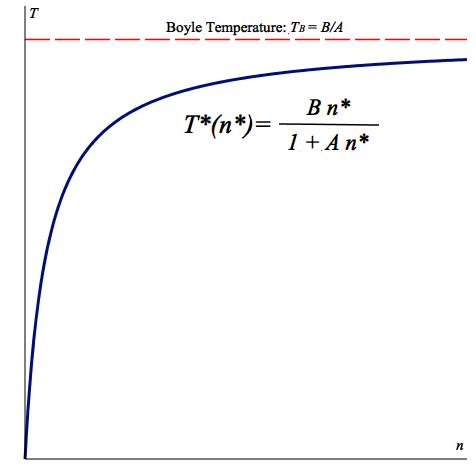}
    \caption{\footnotesize{The curve $T^*(n^*) = Bn^*/(1+An^*)$ is a continuum of equilibrium points for the van der Waals gas. The asymptote is given by the Boyle temperature $T_B = B/A = 10^4$. Clearly, these points exist when $H = 0$ and $p = nT + n^2 T (A - B/T) = 0$ (see the dynamical equations). Vanishing pressure can be achieved only when $F(T) = A - B/T$ is negative, thus the temperature has to be below the Boyle temperature $T_B$. The system will chose one particular equilibrium point $(n^*, H^* = 0, T^*)$ out of this continuum and this will depend entirely on the initial conditions.}}
    \label{figure5}
    \end{figure}

    \noindent
    There is a third invariant manifold --- the second integral $K_3$, defined as the surface $T = 0$. The motion on this surface is given by the equations $\dot{H} = - (3/2) H^2$ (which is the same equation as the equation for $\dot{H}$ on the surface $K_2$) and $\dot{n} = -3 H n$. For initial data $(n_0,H_0)$ at $t=t_0$, the equations again integrate easily and the solutions are
    \b
    \label{t0a}
    H(t) & = & \frac{H_0}{1 + \frac{3H_0}{2} (t - t_0)},\\
    \label{t0b}
    n(t) & = & \frac{n_0}
    {\left[ 1 + \frac{3H_0}{2}  (t - t_0)\right]^{2}}.
    \e
    The long-time asymptotic is again towards  the origin,  $(n,H)\to(0,0),$ where the trajectories converge over infinite time. Similarly, if the equation of state satisfies
    $p(0,T) = 0$, then the critical points $(n^{**}, 0,0)$ are not reachable due to the second integral $K_3$. Thus, a trajectory cannot end anywhere on the invariant plane $T = 0$, except at the origin. \\
    The equilibrium points of interest are the origin and the continuum set $\{ [ n^*, H^* = 0, T^*(n^*) ] \}$ where the pair $n^*$ and $T^*$ are determined by the simultaneous equations $p(n^*, T^*) = 0$ and $I(n^*, T^*) = I(n_0, T_0) = I_1 = $ const.

    \section{Eigenvalues, Eigenvectors and Trajectories. Linearization}

    Consider next the linearised form of the dynamical system near the equilibrium point $[n^\ast, \,\, H^\ast = 0, \,\, T^\ast(n^*)]$:
    \b
    \label{lin1}
    \dot{n} \hskip-.2cm & \equiv & \hskip-.2cm f_1(n, H, T) \! = \!\Bigl(\frac{\partial f_1}{\partial n}\Bigr)^* (n - n^\ast) + \Bigl(\frac{\partial f_1}{\partial H}\Bigr)^*
    (H - H^\ast) + \Bigl(\frac{\partial f_1}{\partial T}\Bigr)^*
    (T - T^\ast) + \ldots, \quad
    \\
    \label{lin2}
    \dot{H} \hskip-.2cm & \equiv & \hskip-.2cm f_2(n, H, T) \! = \! \Bigl(\frac{\partial f_2}{\partial n}\Bigr)^* (n - n^\ast) + \Bigl(\frac{\partial f_2}{\partial H}\Bigr)^*
    (H - H^\ast) + \Bigl(\frac{\partial f_2}{\partial T}\Bigr)^*
    (T - T^\ast) + \ldots, \quad
    \\
    \label{lin3}
    \dot{T} \hskip-.2cm & \equiv & \hskip-.2cm f_3(n, H, T) \! = \! \Bigl(\frac{\partial f_3}{\partial n}\Bigr)^* (n - n^\ast) + \Bigl(\frac{\partial f_3}{\partial H}\Bigr)^*
    (H - H^\ast) + \Bigl(\frac{\partial f_3}{\partial T}\Bigr)^*
    (T - T^\ast) + \ldots. \quad
    \e
    where the stars on the derivatives indicate that they are taken at the equilibrium point $(n^\ast, 0, T^\ast)$. \\
    In matrix form this can be written as:
    \b
    \label{ds1}
    \frac{d}{dt} X(t) = L(n^\ast, H^\ast, T^\ast) \cdot X(t),
    \e
    where:
    \b
    \label{ds2}
    X(t)= \left(
    \begin{array}{c}
    n(t) - n^\ast \cr \cr
    H(t) - H^\ast \cr \cr
    T(t) - T^\ast
    \end{array}
    \right) \qquad
    \mbox{ and } \qquad \,\,
    L(n, H, T) = \left(
    \begin{array}{ccc}
    \frac{\partial f_1}{\partial n} & \frac{\partial f_1}{\partial H} & \frac{\partial f_1}{\partial T}  \cr \cr
    \frac{\partial f_2}{\partial n} & \frac{\partial f_2} {\partial H} & \frac{\partial f_2}{\partial T} \cr \cr
    \frac{\partial f_3}{\partial n} & \frac{\partial f_3}{\partial H} & \frac{\partial f_3}{\partial T}
    \end{array}
    \right).
    \e
    The stability matrix $L(n, H, T)$ at the equilibrium point $[n^\ast, \,\, H^\ast = 0, \,\, T^\ast]$ is:
    \b
    \label{L}
    L^\ast \equiv L(n^\ast, H^\ast, T^\ast)  = \left(
    \begin{array}{ccccc}
    0 & & - 3 n^\ast & & 0
    \cr \cr
    -\frac{1}{2} \left( \frac{\partial p}{\partial n} \right)_{\scriptscriptstyle T}^*  & & 0 & &  -\frac{1}{2} \left( \frac{\partial p}{\partial T} \right)_{\scriptscriptstyle n}^*  \cr \cr
    0 & & -3T^*\left( \frac{\partial p}{\partial \rho} \right)_{\scriptscriptstyle n}^*  & & 0
    \end{array}
    \right).
    \e
    One of the eigenvalues of $L^\ast$ is zero ($\lambda_1 = 0$), while the other two satisfy
    \b
    \label{eigen}
    \lambda^2 = \frac{3}{2} \left[  T^* \frac{\left[ \left(\frac{\partial p}{\partial T}\right)_n^*\right]^2}{\left(\frac{\partial \rho}{\partial T}\right)_n^*}
    + n^* \left( \frac{\partial p}{\partial n} \right)_T^* \right].
    \e
    If the expression on the right-hand side is negative and, say, equal to $- \omega^2$, then $\lambda_{2,3} = \pm i \omega$. If positive and equal to $q^2$, then $\lambda_{2,3} = \pm q$. If zero, then all three eigenvalues are zero\footnote{
    A critical (fixed) point $P$ of a dynamical system is called stable if for any neighbourhood $U(P)$ there is a neighbourhood $V(P)$, such that every trajectory that starts in $V(P)$ does not leave $U(P)$ \cite{crit}. A periodic trajectory $\gamma(t)$ is called orbitally stable if for any neighbourhood $U(\gamma)$ there is a neighbourhood $V[\gamma(0)]$, such that every trajectory that starts in $V[\gamma(0)]$ does not leave $U(\gamma)$ \cite{crit}.} \\
    If the two eigenvalues are imaginary ($\lambda_{2,3} = \pm i \omega$), given the existence of a globally defined first integral $J(n, H, T)$, and if the hyper-surface, defined by the other first integral $I(n,T)$, is well behaved around $(n^*, 0, T^*)$ (which is to be expected from the physical context), then there are periodic trajectories which are always confined by the integral surfaces and are orbitally stable. They are orbiting around the critical point $(n^*, 0, T^*)$ and this critical point is stable. If the eigenvalues are real, $\lambda_{2,3} = \pm q$  (they are with opposite signs), then the critical point $(n^*, 0, T^*)$ is unstable. \\
    The eigenvectors $u_i$, corresponding to eigenvalues $\lambda_i, \,\, i=1,2,3, \,$ are:
    \b
    u_1 =
    \left(
    \begin{array}{c}
    1  \cr
    0 \cr
    \left( \frac{\partial T}{\partial n} \right)^*_p
    \end{array}
    \right), \,\,
    u_{2,3} =
    \left(
    \begin{array}{c}
    -3 n^\ast \cr
    \pm i \omega \cr
    -3 T^* \left( \frac{\partial p }{\partial \rho} \right)^*_n
    \end{array} \right)
    \,\, \mbox{or}   \,\,
    u_{2,3} =
    \left(
    \begin{array}{c}
    -3 n^\ast \cr
    \pm q \cr
    -3 T^* \left( \frac{\partial p}{\partial \rho} \right)^*_n
    \end{array}
    \right),
    \e
    where the choice $\pm i \omega$ or $\pm q$ depends on the sign of the right-hand side of (\ref{eigen}). \\
    It is interesting to study the closed trajectories which correspond to $\lambda_{2,3} = \pm i \omega$ (see Figure 6). \\
    Replacing $L(n^\ast, H^\ast, T^\ast)$ in (\ref{ds1}) with $M \Lambda M^{-1}$, where $M$ is the matrix whose columns are the eigenvectors $u_{1,2,3}$ and $\Lambda$ is the diagonal matrix with the eigenvalues $\lambda_{1,2,3}$ along its main diagonal, multiplying (\ref{ds1}) from the left with $M^{-1}$, noting that $M^{-1} = $ const, and introducing the column-vector $Z(t) = [z_1(t), z_2(t), z_3(t)] = M^{-1} X(t)$, allows to write the dynamical system in diagonalised form: $\dot{z}_i(t) = \lambda_i z_i (t)$ with $i = 1,2,3. $ Thus $z_1 = \alpha_1 = $ const, $z_{2,3} (t) = \alpha_{2,3} e^{\pm i \omega t}, $ where $\alpha_{2,3} = $ const. \\
    Using $X(t) = M Z(t)$ and requesting reality for the dynamical variables --- which, in turn, necessitates that $z_2(t) = \bar{z}_3(t)$, that is, $\alpha_2 = \bar{\alpha}_3 \equiv \alpha$ --- allows to find:
    \b
    \label{a1}
    n(t) - n^\ast & = & \alpha_1  - 3 n^* [z_2(t) + z_3(t)] = \alpha_1 - 6 \alpha n^\ast \cos \omega t,
    \\
    \label{a2}
    H(t) - 0 \phantom{e} & = & i \omega
    [z_2(t) - z_3(t)] = 2 \alpha \omega \sin \omega t,
    \\
    \label{a3}
    T(t) - T^\ast & = & \alpha_1 \left(\frac{\partial T}{\partial n} \right)^*_p - 3 T^* \left(\frac{\partial p}{\partial \rho} \right)^*_n [z_2(t) + z_3(t)] \nonumber \\
    & = & \alpha_1 \left(\frac{\partial T}{\partial n} \right)^*_p - 6 \alpha T^* \left(\frac{\partial p}{\partial \rho} \right)^*_n \cos \omega t.
    \e
    One can immediately deduce from here that:
    \b
    \label{plane}
    [n(t) - n^\ast] - \frac{n^\ast}{T^* \left( \frac{\partial p}{\partial \rho}\right)_n^*} [T(t) - T^{\ast}]
    = \alpha_1 \left[ 1 - \frac{n^*}{T^*} \frac{\left( \frac{\partial T}{\partial n} \right)^*_p}{\left(\frac{\partial p}{\partial \rho} \right)^*_n} \right].
    \e
    The right-hand side is a constant, which is not necessarily small. In the linear approximation however, $n$ is very close to $n^*$ and $T$ is very close to $T^*$. Thus, this constant must be set equal to zero, namely, either $\alpha_1$ should be zero or the bracketed term on the right-hand side should be zero. The latter cannot be taken as zero because this would impose an extra relationship on $p$, $n$, $\rho$, and $T$. Thus one must have $\alpha_1$ equal to zero. \\
    Without loss of generality, $\alpha$ can be taken as positive.

    \begin{figure}[!ht]
    \centering
    \subfloat[\scriptsize The trajectory is an almost flat "ellipse".]{\label{F17}\includegraphics[height=3.6cm, width=0.28\textwidth]{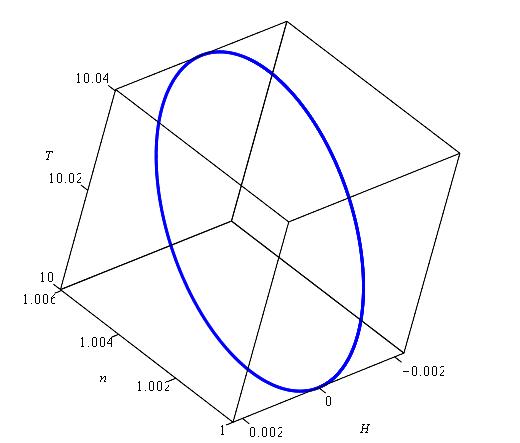}}
    \,
    \subfloat[\scriptsize Projection of the trajectory onto the $n$--$T$ plane.]{\label{F18}\includegraphics[height=3.3cm,width=0.22\textwidth]{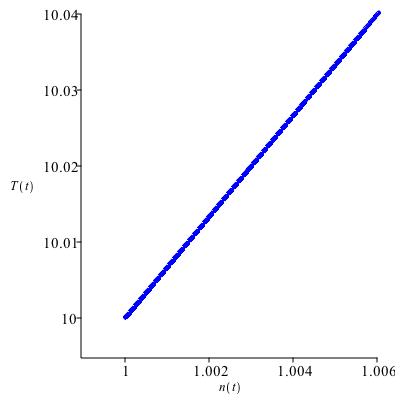}}
    \,
    \subfloat[\scriptsize Projection of the trajectory onto the $n$--$H$ plane.]{\label{F19}\includegraphics[height=3.3cm,width=0.24\textwidth]{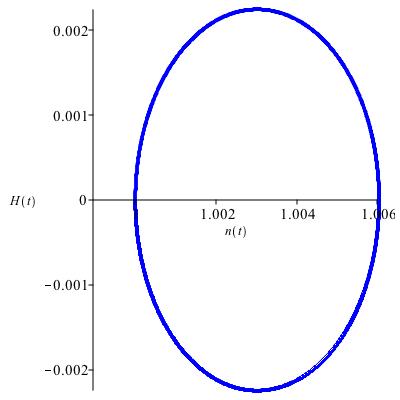}}
    \,
    \subfloat[\scriptsize Projection of the trajectory onto the $T$--$H$ plane.]{\label{F20}\includegraphics[height=3.3cm,width=0.22\textwidth]{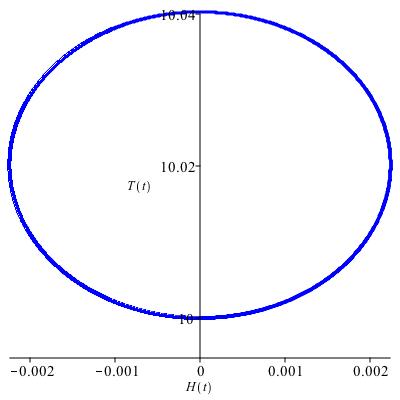}}
    \caption{\footnotesize{A trajectory in the linearly-approximated van der Waals phase model. The initial conditions are: $n_0 = 1$, $H_0 = 0$, and $T_0 = 20$ --- same as the innermost closed curve in Figure 3(a).}}
    \label{figure6}
    \end{figure}

    \noindent
    The trajectory in the three-dimensional phase space is a closed curve and the equation of its projection (in the linear approximation) onto the $n$--$H$ plane is an ellipse (see Figure 6c):
    \b
    \label{ell1}
    \left[ \frac{n(t) - n^\ast}{6\alpha n^\ast}\right]^2+\left( \frac{H}{2\alpha \omega}\right)^2 = 1,
    \e
    while the equation of its projection onto the $T$--$H$ plane is another ellipse (Figure 6d):
    \b
    \label{ell2}
    \left[ \frac{T - T^\ast}{6\alpha T^* \left(\frac{\partial p}{\partial \rho} \right)^*_n}\right]^2+\left( \frac{H}{2\alpha \omega}\right)^2 = 1.
    \e
    These are two ellipses with semi-axes proportional to $\alpha$ in each of the cases. The constant $\alpha$, itself, is fixed by the initial conditions. \\
    Equation (\ref{plane}) is the equation of a plane and this plane is the tangent plane at point $(\rho^\ast,0,T^\ast)$ (the centre) to the surface along which the dynamical system evolves in the three-dimensional phase space. This  plane (\ref{plane}), intersected with the plane $H = 0$, yields the straight line (Figure 6b):
    \b
    \label{line}
    T(t) - T^* = \frac{T^*}{n^*} \left( \frac{\partial p}{\partial \rho}\right)_n^* [n(t) - n^*]
    \e
    in the linear approximation. \\
    To illustrate the above linear approximation with an example, start with the eigenvalues for a real virial gas system with equation of state given by (\ref{eos}). These are $\lambda_1 = 0$ and $\lambda_{2,3}$ satisfying:
    \b
    \label{eigenvir}
    \lambda^2 =  \frac{n^{*^3} T^{*^3} F'{^2}(T^*)}{1 - \frac{2}{3} n^* T^* [ 2 F'(T^*) + T^* F''(T^*)]} - \frac{3}{2} n^* T^*.
    \e
    If one again considers a van der Waals gas [for which $F(T) = A - B/T$], then:
    \b
    \label{eigenvdw}
    \lambda^2 = \frac{B n^{*^2}}{1 + A n^*} \left[ (1+A n^*)^2 -\frac{3}{2} \right].
    \e
    These are purely imaginary if $n^* < (\sqrt{3/2} - 1) / A$. Given that $n^* = T^* / (B - A T^*)$ for the van der Waals gas, one gets that closed trajectories exist for $T^* < (1 - \sqrt{3/2}) T_B \approx 0.1835 \, T_B$ (recall that $T_B = B/A$). If this is the case, then the trajectory in the linear approximation is an ellipse and the projection of this ellipse onto the the $n$--$H$ plane is the same as (\ref{ell1}), while its projection onto the $T$--$H$ plane is
     \b
    \left( \frac{T - T^\ast}{6\alpha B n^*}\right)^2+\left( \frac{H}{2\alpha \omega}\right)^2 = 1.
    \e
    From these two, or from (\ref{line}), one gets that the projection of the trajectory onto the $n$--$T$ plane is given by
    \b
    T(t) - T^* = \frac{2}{3} \, B [n(t) - n^*].
    \e

    \section{Periods of Inflation and Deflation for the Linearized Theory}

    To achieve solution describing inflation [by definition, inflation is equivalent to $\ddot{a}(t) > 0$ and $\dot{a}(t) > 0$], it is obvious from (\ref{hash}) that negative pressure (\ref{eos}) is needed. Inflation occurs when
    \b
    \label{cond}
    \frac{\ddot{a}}{a} = \dot{H} + H^2 = -\frac{1}{2} H^2 - \frac{1}{2} p > 0
    \e
    since the scale factor $a(t)$ is always strictly positive. That is, when $p < - H^2$. \\
    The inflation parameter $\varepsilon_{\mbox{\tiny I}}$, given by:
    \b
    \varepsilon_{\mbox{\tiny I}} = - \frac{\dot{H}(t)}{H^2(t)}
    = \frac{d}{dt}\frac{1}{H(t)}
    = -\frac{d \ln H(t)}{dN(t)},
    \e
    must be less than 1. Here $dN(t) = d \ln a(t) = H(t) dt$ measures the number $N$ of $e$-folds of inflationary expansion. \\
    For the most general gas which admits negative pressure, let $G(n, T)$ denote the region(s) in the $n$--$T$ plane for which $p(n,T) < - H^2$.
    If inflation occurs, it will be over $G(n,T)$. \\
    To analyse the situation in the linear approximation (Figure 6), it is most convenient to firstly find the scale factor $a(t)$. When $H(t)$ is given by (\ref{a2}), i.e. $H(t) = 2 \alpha \omega \sin \omega t$ (purely imaginary eigenvalues):
    \b
    a(t)  & = & a_0 e^{-2\alpha \cos \omega t} \\
    \dot{a}(t) & = & 2 \alpha \omega a(t) \sin \omega t, \\
    \ddot{a}(t) & = & 2\alpha \omega^2 a(t) (2\alpha \sin^2 \omega t + \cos \omega t ) \nonumber \\
    & = & 2\alpha \omega^2 a(t)
    \left[ 2\alpha\left[ 1+\frac{1}{(4\alpha)^2}\right] -2\alpha \left(\cos \omega t -\frac{1}{4\alpha}\right)^2 \right].
    \e
    Recalling that $\alpha > 0$, positivity of $\dot{a}(t)$ means that $0 < \omega t < \pi$, while positivity of $\ddot{a}(t)$ means:
    \b
    \frac{1}{4\alpha} -
    \sqrt{1+\frac{1}{(4\alpha)^2}}
    < \cos \omega t
    < \frac{1}{4\alpha} + \sqrt{1+\frac{1}{(4\alpha)^2}}.
    \e
    Note that $1/(4 \alpha) + [1 + (4\alpha)^{-2}]^{1/2}$ is greater than 1 for all $\alpha$, thus the second inequality is always satisfied. \\
    On the other hand, $-1 < 1/(4\alpha) - [1 + (4\alpha)^{-2}]^{1/2}  < 0$ for all $\alpha$. It is convenient to introduce
    $t_\alpha$ via
    \b
    \cos \omega t _{\alpha}= \frac{1}{4\alpha} -
    \sqrt{1+\frac{1}{(4\alpha)^2}},
    \e
    so that $\pi/2 < \omega t_\alpha < \pi.$
    Therefore, inflation occurs for times $t$ satisfying
    \b
    \label{infla}
    0 < \omega t < \omega t_{\alpha}.
    \e
    The upper limit depends on the initial conditions via $\alpha$. Therefore, inflation occurs for $\omega t$ between 0 and, at least, $\pi / 2$.\\
    One should note that the model also includes deflation --- occurring for $\omega t$ between $\pi$ and, at least $3 \pi / 2$ (this can be seen in analogous manner). \\
    To see how much inflation is generated, consider the number $N$ of $e$-folds of inflationary expansion: $dN(t) = H(t) dt$. This integrates easily to give $N = 2 \alpha$ and this depends on the initial data through $\alpha$. \\
    If the eigenvalues are real, then the trajectories are open curves and $H(t) = 2 \alpha q \sinh q t$. Thus:
    \b
    \label{end1}
    a(t)  & = & a_0 e^{2 \alpha \cosh q t} \\
    \label{end2}
    \dot{a}(t) & = & 2 \alpha q a(t) \sinh q t, \\
    \label{end3}
    \ddot{a}(t) & = & 2 \alpha q^2 a(t) (2 \alpha \sinh^2 q t + \cosh q t ).
    \e
    Since $\alpha$, $q$, and time $t$ are all positive, then $\dot{a}(t)$ will also be positive. The acceleration $\ddot{a}(t)$ is also positive. Therefore, a ``fly-by'' near an unstable critical point (i.e. in the neighbourhood of the unstable critical point, defined as the region where the linear approximation is applicable) is always characterised by inflation. \\
    For a van der Waals gas at very high densities and temperatures, the leading term in $\dot{H}$ from (\ref{dve''}) is $-\frac{A}{2}n^2T$, while the leading term in $H^2$ from (\ref{i2-vdw}) is $\frac{3}{2}n T$. Thus
    \b
    \varepsilon \simeq \frac{An}{3}\to \infty
    \e
    which disagrees with the inflation condition. Therefore, near the blow-up time inflation is not observed.

    \section{Hamiltonian Formulation}

    The two integrals of motion completely determine the global behaviour of the full system --- see Figure 2. In the case of van der Waals gas, using $J_{vdw}(n, H, T)$ from (\ref{i2-vdw}) and with the help of (\ref{tova}), allows to exclude the temperature from the picture: $T(n)=Dn^{2/3}\exp(2An/3)$. The dynamics is then governed by:
    \b
    \label{dve-edno}
    \dot{n} & = & -3Hn,  \\
    \label{dve-dve}
    \dot{H} & = & -\frac{3}{2}H^2 -\frac{1}{2} p(n),
    \e
    where $p(n)$ is the pressure expressed in terms of the number density $n$ and for the van der Waals gas it is given by:
    \b
    \label{pe2}
    p(n)= - B n^2 + D (1 + A n) \, n^{\frac{5}{3}} \, e^{\frac{2An}{3}}.
    \e
    The critical points of the underlying two-dimensional system occur where the Hubble parameter and the pressure $p(n)$ vanish, namely, these are the points $(n^\ast = 0, H^\ast = 0)$ and $(n^{**}, H^{**} = 0)$. The zero $n^{**}$ of (\ref{pe2}) depends on the van der Waals gas parameters $A$ and $B$ and on the initial conditions via $D$. \\
    For the most general equation of state, solving (\ref{simet}) leads to the appearance of the implicit integral $I(n,T) = $ const and, again, $T = T(n)$ can be implicitly excluded, resulting in the same dynamical system: (\ref{dve-edno}) and (\ref{dve-dve}), with the relevant $p(n)$. \\
    Introducing new variables $u(n) = 2 / (3 \sqrt{n})$ and $v = H / \sqrt{n}$, the system becomes
    \b
    \dot{u} & = & v, \\
    \dot{v} & = & \varphi(u),
    \e
    where $\varphi(u) = - p(n) / (2\sqrt{n}).$ \\
    It can be written in terms of the canonical variables $u$ and $v$ with Hamiltonian (conserved quantity)
    \b
    \mathscr{H}(u,v)= \frac{1}{2}v ^2 -\int\limits_{0}^{u} \varphi(\widetilde{u}) d \widetilde{u}
    \e
    and dynamics:
    \b
    \dot{u}& = & \frac{\partial \mathscr{H}}{\partial v} =  v, \\
    \dot{v}& = & -\frac{\partial \mathscr{H}}{\partial u} = \varphi(u).
    \e
    Even in the case of a most general equation of state, the underlying two-dimensional system is Hamiltonian. As it is well known, for Hamiltonian systems the only allowed critical points are centres and saddles. Therefore, in the case of imaginary eigenvalues, one is dealing with a centre-type behaviour. \\
    The system corresponds to one-dimensional motion of a particle in a potential field $\mathscr{V}(u) = - \int_{0}^{u} \varphi(\widetilde{u}) d \widetilde{u}$ --- see \cite{A} and {\cite{vilasi}. The centre-type oscillations are near the local minima of $\mathscr{V}(u)$. \\
    It is not difficult to check that the centres of the canonical system transform back to centres of the original system (\ref{dve-edno}), (\ref{dve-dve}) in terms of $n$ and $H$.

    \section{Discussion}

    The type of the dynamical behaviour of the system is influenced by two groups of parameters --- the initial data of the system $(n_0, H_0, T_0)$ and the gas parameters from the equation of state. The integral curve is an intersection of the level sets of two global first integrals and the possible integral curves are classified in dependence on the aforementioned parameters.  A distinct feature of the system are the stable periodic orbits around the critical point $[n^*,0,T^*(n^*)]$. They occur in the case of phantom dust and for imaginary eigenvalues --- see (\ref{eigen}) and (\ref{eigenvir}) for the case of virial gas. Such scenario depends on the gas parameters as well, which can be seen from formula (\ref{eigenvdw}) for the van der Waals gas. Inflationary and deflationary periods in the cyclic regime are observed (\ref{infla}). Alternatively, when the eigenvalues at $[n^*,0,T^*(n^*)]$ are real (positive and negative), there are no closed orbits, but all ``fly-by'' trajectories near $[n^*,0,T^*(n^*)]$ are characterised by inflation  (\ref{end1})--(\ref{end3}). \\
    In the case of positive dust density, the trajectories end up asymptotically either at $(0,0,0)$ or are unbounded (Big Chrunch). The special explicit solutions in the invariant planes $n=0$ (\ref{n0a})--(\ref{n0b}) and $T=0$ (\ref{t0a})--(\ref{t0b}), as well as the exact van der Waals solution in this asymptotic regime (\ref{assy}), indicate that the trajectories terminating at the origin $(n,H,T) \to (0, 0, 0)$ behave asymptotically as negative powers of $t$ and reach the origin in infinite time. This excludes possible inflation near the origin. The unbounded van der Waals trajectories (Figure 3b) represent a blow-up in finite time (Big Crunch) and are not inflatory.

    \end{document}